\documentclass{IEEEtran}
\IEEEoverridecommandlockouts
\usepackage{cite}
\usepackage{amsmath,amssymb,amsfonts}
\usepackage{algorithmic}
\usepackage{graphicx}
\usepackage{textcomp}
\usepackage{xcolor}
\usepackage{setspace}
\usepackage{mathtools}

\def\BibTeX{{\rm B\kern-.05em{\sc i\kern-.025em b}\kern-.08emT\kern-.1667em\lower.7ex\hbox{E}\kern-.125emX}}
\usepackage{subfigure}

\usepackage{enumitem}

\title{Exact Performance Analysis of THz Link Under	Transceiver Hardware Impairments}

\author{Pranay Bhardwaj,~\IEEEmembership{Graduate Student Member,~IEEE} and S.~M.~ Zafaruddin,~\IEEEmembership{Senior Member,~IEEE}
	\thanks{Pranay Bhardwaj (p20200026@pilani.bits-pilani.ac.in) and S.~M.~Zafaruddin (syed.zafaruddin@pilani.bits-pilani.ac.in)  are with the Department of Electrical and Electronics Engineering, Birla Institute of Technology and Science, Pilani, Pilani-333031, Rajasthan, India.}
	\thanks{This work was supported	in part by the Science and Engineering Research Board (SERB), Department of Science and Technology (DST), Government of India, through the Mathematical Research Impact Centric Support (MATRICS) scheme under Grant MTR/2021/000890.}
}

\begin{document}
	\maketitle 
	\begin{abstract}
Transceiver hardware impairment (THI) is inevitable for high-date rate terahertz (THz) communication. Existing statistical analysis either neglects THI's effect or provides approximate results when analyzing the performance of the THz system combined with channel fading and antenna misalignment. In this paper, we develop exact analytical expressions for the average signal-to-noise ratio (SNR), ergodic capacity, and average bit-error-rate (BER) performance of a THz wireless link under the combined effect of $\alpha$-$\mu$ fading channel, zero-boresight pointing errors, and the Gaussian distributed THI. We also derive asymptotic expressions for the outage probability and average BER, which  shows that the diversity order of the THz link is independent of THI's parameters. Simulations validate the derived analytical results and demonstrate the impact of the THI parameters on the THz performance.
	\end{abstract}				
	
	\section{Introduction}
Terahertz (THz) wireless technology can provide ultra-high data-rate transmission for short link distances using enormous channel bandwidth available in the THz spectrum \cite{Akyildiz_2020,Dang_2020_nature}. However, transceiver hardware impairment (THI) inevitably degrades the performance of  THz communication at a high data rate. The THI is caused by imperfections such as an imbalance in in-phase and quadrature-phase signal components, phase noise, and non-linearities in the power amplifier \cite{Schenk2008}. In addition to the THI, the THz link is susceptible to channel impairments such as the deterministic path-loss and statistical effects from short-term fading and pointing errors \cite{Boulogeorgos_Error}. Accurate performance analysis of the THz link under the combined effect of the THI and channel impairments is desirable for a better design and deployment assessment of THz wireless systems.

Recently, there has been a surge in analyzing the performance of THz wireless transmission for various network configurations \cite{Boulogeorgos_Error,Boulogeorgos_2020_THz_THz,Pranay_2021_TVT,Li_2021_THz_AF,Bhardwaj2022_multihop}. These research employ novel approaches to provide an exact analysis of THz systems since the statistical characterization of the channel impairments in the THz link is complicated due to the generalized $\alpha$-$\mu$ distribution for the short-term fading with a zero-boresight statistical model for pointing errors. However, the impact of THI was ignored as a first approximation in analyzing the performance limits of THz systems \cite{Boulogeorgos_Error,Boulogeorgos_2020_THz_THz,Pranay_2021_TVT,Li_2021_THz_AF,Bhardwaj2022_multihop}. Nevertheless, the authors in \cite{Boulogeorgos_Analytical, Boulogeorgos2022_rain, Du2022_FTR_THz_RIS,Chapala2021_letters} considered the impact of THI along with channel impairments for THz systems. Boulogeorgos \emph{et. al} \cite{Boulogeorgos_Analytical, Boulogeorgos2022_rain} provided an exact expression of the outage probability but an upper bound on the ergodic capacity under the combined effect of THI and channel impairments. The authors in \cite{Du2022_FTR_THz_RIS} derived an upper bound of the ergodic capacity for reconfigurable intelligent surfaces (RIS) assisted THz wireless system. Even for simpler radio-frequency fading models, approximations and upper bounds on the performance are available under the statistical effect of the THI \cite{Bjornson2013_HW_dualhop,Balti2018_HW_FSORF}. Thus, there is a gap in the study. The statistical characterization of the signal-to-noise ratio (SNR) of a THz link under the combined effect of the Gaussian THI, channel fading, and random pointing errors becomes complicated, prohibiting the  direct application of conventional methods to derive exact analytical expressions for various performance metrics such as average SNR, ergodic capacity, and average bit-error-rate (BER). 
 
This article provides a novel approach to an exact analysis of various performance metrics of a THz wireless link under the combined effect of $\alpha$-$\mu$ fading channel, zero-boresight pointing errors, and the Gaussian distributed THI. We present analytical expressions for the average SNR, ergodic capacity, and average BER performance in terms of Fox's H-function. We also derive asymptotic expressions for the outage probability and average BER to show that the diversity order of the THz link is independent of THI's parameters. Simulations validate the derived analytical results and demonstrate the impact of the THI parameters on the THz performance under various scenarios.

\emph{Notations:} The symbol $\mathcal{CN}(\cdot, \cdot)$ denotes the complex normal distribution. $\Gamma(u)= \int\limits_{0}^{\infty}t^{u-1} e^{-t}dt$ represent the Gamma function, while $ \gamma_{in}(u,z)=\int_{0}^{z}s^{u-1}e^{-s}ds$ and $\Gamma(u,z)=\int_{z}^{\infty}s^{u-1}e^{-s}ds$ indicate the lower and upper incomplete Gamma functions, respectively. $G_{p,q}^{m,n}\big(.|.\big)$ symbolizes the Meijer's G-function whereas $ H_{p,q}^{m,n}\big(.|.\big) $ represents the Fox's H-function \cite{Mathai_2010}.
\section{System Model}\label{sec:system_model}
	Consider a THz wireless communication link of distance $l$  with  the received signal  $y$ as
	\begin{eqnarray}\label{model_smpl}
		y = h_l   h  (s+w_{t}) + w_{r} + w
	\end{eqnarray}
	where $h_l $ is the path gain from the source to the destination, $h$ is the channel fading coefficient for both short-term and pointing errors, $s$ is the transmitted signal with power $P$, and $w$ is the additive Gaussian noise with variance $\sigma_w^2$. Further, $w_t$ and $w_r$ denote the components for hardware impairments modeled statistically using Gaussian distribution as $w_{t}\sim\mathcal{CN}(0,k_{t}^{2}P)$, and $w_{r}\sim\mathcal{CN}(0,k_{r}^{2}P\lvert h_l  h \rvert^{2})$ with factor $k_{t}$ and $k_{r}$ characterizing the extent of hardware imperfections in the transmitter and receiver, respectively.	
	
	The deterministic parametric model for the path gain $h_l$ consisting of molecular absorption losses at THz bands is available in various literature \cite{Boulogeorgos_Analytical}. The combined effect of short-term fading (distributed according to the $\alpha$-$\mu$ \cite{Papasotiriou2021_scientific_report}) and antenna misalignment errors (zero-boresight model \cite {Farid2007}) is statistically modeled using the PDF as \cite{Boulogeorgos_Error,Pranay_2021_TVT,Li_2021_THz_AF,Bhardwaj2022_multihop}
	\begin{eqnarray}\label{hi_model_PDF}
		f_{\lvert h_{} \rvert} (x) = \psi x^{\phi-1} \Gamma \bigg(\mu-{\frac {\phi}{\alpha},\zeta x^{\alpha}}\bigg)
	\end{eqnarray}
	where $\psi =  \frac {\phi S^{-\phi} \mu^{\frac{\phi}{\alpha}}}{\Omega^{\phi} \Gamma (\mu)}$ ,and $\zeta = \frac {\mu S^{-\alpha}}{\Omega^{\alpha}} $. Here, $\alpha$ and $\mu$ are fading channel parameters while $S$ and $\phi$ determine the extent of  pointing errors. The parameter $ S $ represents the effective received power with perfect antenna alignment, and $\phi$ signifies the ratio between equivalent beam radius and standard deviation of the pointing error displacement at the receiver. Note that the statistical model in \eqref{hi_model_PDF} has been extensively studied for various network configurations for THZ transmission without the effect of THI. 
	The CDF for the THz link is derived in \cite{Pranay_2021_TVT}:
\begin{flalign}	\label{hi_model_CDF}
	&F_{|h|}(x)=  \frac{\psi}{\phi} \bigg[ \gamma_{in}(\mu,\zeta x^{\alpha})   + \zeta^{\frac{\phi}{\alpha}} x^{\phi} \Gamma\Big(\mu-\frac{\phi}{\alpha},\zeta x^{\alpha}\Big) \bigg]
\end{flalign}
Using \eqref{model_smpl}, the resultant  SNR with the THI can be expressed as 
\begin{eqnarray}\label{eq:thi_snr}
\gamma = \frac{\gamma_0 {\lvert h \rvert} ^2} {k^2 \gamma_0 {\lvert h \rvert}^2 + 1}
\end{eqnarray}
where $\gamma_0 = \frac {P{\lvert h_l \rvert}^2}{\sigma_w^2}$ and $k^2 = {k_t}^2 + {k_r}^2$. 

In what follows, we use \eqref{eq:thi_snr} to  analyze the performance of the THz link to assess the impact of the THI in addition to other channel impairments such as path loss, fading, and misalignment errors. It should be mentioned that the approach of our analysis can also be carried out for different  fading channels and pointing error model \cite{Dabiri2022}.
 
	 \section{Statistical Performance Analysis}
	 In this section, we provide an exact statistical analysis of the outage probability, average BER, ergodic capacity, and the $n$-th moment of SNR for THz wireless links under the THI. We also derive asymptotic expressions at high SNR for the outage probability and average BER, providing engineering insight into the effect of system parameters. 
	 
	To analyze the performance of the THz link, we require density and distribution functions of the SNR. Using \eqref{hi_model_PDF} and \eqref{hi_model_CDF} in \eqref{eq:thi_snr} with the standard transformation of random variables \cite{papoulis_2002}, the PDF and CDF of the SNR for a THz link under the combined effect of short-term fading, pointing errors, and THI can be derived as 
	 	 \begin{eqnarray} \label{simplified_pdf}	 	
	 f_\gamma(\gamma) = {\frac{\psi  \gamma^{\frac{\phi}{2}-1}}{2\gamma_0^\frac{\phi}{2} (1-\gamma k^2)^\frac{\phi}{2}}}  \Gamma \bigg(\mu - \frac{\phi}{\alpha},\zeta{\Big(\sqrt{\frac{\gamma}{\gamma_0(1-\gamma k^2)}}\Big)}^{\alpha}\bigg)	
	 \end{eqnarray}
	 \begin{flalign} \label{cdf_model}
	 &F_\gamma(\gamma) = \frac{\psi {\zeta}^{\frac{-\phi}{\alpha}}}{\phi}\Bigg[\gamma_{in}\bigg(\mu,\zeta{\Big(\sqrt{\frac{\gamma}{\gamma_0(1-\gamma k^2)}}\Big)}^{\alpha}\bigg) \nonumber \\  & + {\zeta}^{\frac{\phi}{\alpha}} \bigg(\sqrt{\frac{\gamma}{\gamma_0(1-\gamma k^2)}}\bigg)^{\phi}  \Gamma \bigg(\mu - {\frac{\phi}{\alpha},\zeta{\Big(\sqrt{\frac{\gamma}{\gamma_0(1-\gamma k^2)}}\Big)}^{\alpha}}\bigg)\Bigg]
	 \end{flalign}
	 \subsection{Outage Probability}
	 The outage probability is an important performance metric depicting the probability that a communication system under the fading channel goes into an outage when the instantaneous SNR drops below a predefined threshold level, i.e., $P_{\rm out} = P(\gamma<\gamma_{\rm th})$. An exact expression for the outage probability can be obtained using  $\gamma = \gamma_{\rm th}$ in the CDF \eqref{cdf_model} as $P_{\rm out} =F_\gamma(\gamma_{\rm th})$.

 Using the series expansion of incomplete gamma function \cite{Jameson2016} as $\gamma_0\to \infty$, an asymptotic expression for the outage probability in a high SNR regime can be obtained as
	\begin{flalign} \label{eq:outage_asymptotic_high_snr}
		{P_{\rm out}^{\infty}} = &  \frac{S_0^{-\alpha\mu}\mu^{\mu} }{\Gamma(\mu) \mu}  \bigg(\frac{\gamma_{\rm th}}{\gamma_{0}(1-k^2)}\bigg)^{\frac{\alpha\mu}{2}} \nonumber \\  + & \frac{S_0^{-\phi}\mu^{\frac{\phi}{\alpha}} \Gamma(C)}{\Gamma(\mu)} \bigg(\frac{\gamma_{\rm th}}{\gamma_{0}(1-k^2)}\bigg)^{\frac{\phi}{2}}
\end{flalign}

Capitalizing the asymptotic expression in \eqref{eq:outage_asymptotic_high_snr} with the exponents of $\gamma_{0}$, the diversity order can be obtained as
 \begin{flalign} \label{eq:diversity_order_outage}
 	DO = \min\Bigl\{\frac{\alpha\mu}{2}, \frac{\phi}{2}\Bigr\}.
 \end{flalign}

It can be observed from \eqref{eq:diversity_order_outage} that the THI parameter $k$ does not change the diversity order of the system but has an effect on the coding gain.

\subsection{Average BER}
Noting the range  $0<\gamma<\frac{1}{k^2}$, a general formulation for the average BER for the THz link with the SNR in \eqref{eq:thi_snr} using the CDF $F_\gamma(\gamma)$ is given by
\begin{eqnarray}\label{BER}
	\bar{P}_{e} = \frac{q^p}{2\Gamma(p)} \int_{0}^{\frac{1}{k^2}} e^{-q\gamma} \gamma^{p-1} F_{\gamma} (\gamma) d\gamma
\end{eqnarray}
where the set $\{p,q\}$ specifies a particular set of modulation scheme. We substitute \eqref{cdf_model} in \eqref{BER} to get
\begin{flalign}\label{eq:BER_1}
	&\bar{P}_{e} = \frac{\psi {\zeta}^{\frac{-\phi}{\alpha}} q^p}{\phi2\Gamma(p)} \int_{0}^{\frac{1}{k^2}} e^{-q\gamma} \gamma^{p-1} \Bigg[\gamma_{in}\bigg(\mu,\zeta{\Big(\sqrt{\frac{\gamma}{\gamma_0(1-\gamma k^2)}}\Big)}^{\alpha}\bigg) \nonumber \\  & + {\zeta}^{\frac{\phi}{\alpha}} \bigg(\sqrt{\frac{\gamma}{\gamma_0(1-\gamma k^2)}}\bigg)^{\phi}  \Gamma \bigg(\mu - {\frac{\phi}{\alpha},\zeta{\Big(\sqrt{\frac{\gamma}{\gamma_0(1-\gamma k^2)}}\Big)}^{\alpha}}\bigg)\Bigg] d\gamma
\end{flalign}
A direct application of integral identities is not applicable for \eqref{eq:BER_1} considering the finite limit of integration with integrand involving incomplete Gamma function with the ratio of variable in its argument. Thus, we make a judicially chosen substitution $({\frac{1}{\gamma k^2}-1 })^{\frac{\alpha}{2}}=t$ in \eqref{eq:BER_1} to get
\begin{flalign}\label{eq:zaf1}
\bar{P}_{e} =& - \frac{a q^p}{2\Gamma(p)} \Bigg[\frac{\psi \zeta^\frac{-\phi}{\alpha}}{\phi} \int_{0}^{\infty} e^{\frac{-q}{k^2(t^a+1)}} \gamma_{in}(\mu,B t) \nonumber \\ \times &  \bigg(\frac{1}{k^2 (t^a+1)^2}\bigg)^{(p-3)}  t^{(a-1)} dt +\frac{\zeta^{\frac{\phi}{\alpha}}}{{\gamma_0}^\frac{\phi}{2} k^{\phi}} \int_{0}^{\infty} t^{-b} \Gamma(C,Bt) \nonumber \\ \times & e^{\frac{-q}{k^2(t^a+1)}}  \bigg(\frac{1}{k^2 (t^a+1)^2}\bigg)^{(p-3)}   t^{(a-1)} dt\Bigg]
\end{flalign}
where  $C=\mu-\frac{\phi}{\alpha}$, $B = \frac{\zeta}{{\gamma_0}^\frac{\alpha}{2} k^{\alpha}}$, $ a = \frac{2}{\alpha} $, and  $b=\frac{\phi}{\alpha}$. Since there is no closed-form representation of \eqref{eq:zaf1}, even with Meijer's G approach, we use another substitution $t = (z-1)^{\frac{1}{a}}$ and utilize the Mellin–Barnes type integral form of Meijer's G-function for the exponential and incomplete Gamma functions to express \eqref{eq:zaf1} as
\begin{flalign}\label{meiger_def_BER_1}
	\bar{P}_{e} = & -\frac{q^p}{2\Gamma(p)} \Bigg[ \frac{\psi \zeta^\frac{-\phi}{\alpha}}{\phi} \frac{1}{(2\pi i)^3}\int_{L_1}^{} \int_{L_2}^{}\int_{L_3}{} \frac{\Gamma(\mu-s_1)\Gamma(1+s_1)}{\Gamma(1+s_1)}            \nonumber \\ \times & \Gamma(-s_2) \Gamma(p-3+s_2) \Gamma(-s_3) ds_1 ds_2 ds_3 \times I_1     \nonumber \\ + &   \frac{\zeta^{\frac{\phi}{\alpha}} }{{\gamma_0}^\frac{\phi}{2} k^{\phi}} \frac{1}{(2\pi i)^3} \int_{L_1}^{} \int_{L_2}^{}\int_{L_3}{}  \frac{\Gamma(C-s_1)\Gamma(-s_1)}{\Gamma(1-s_1)} \Gamma(-s_2) \nonumber \\ \times &  \Gamma(p-3+s_2)  \Gamma(-s_3)  ds_1 ds_2 ds_3 \times I_2   ]
\end{flalign} 
where $I_1$ and $I_2$ are expressed in terms of Gamma functions using the identity \cite[3.191.2]{Gradshteyn}: 
\begin{flalign}\label{inner_integral_BER_2}
	I_1 =& \int_{1}^{\infty} (B(z-1))^{\frac{s_1}{a}} {(z-1)}^{s_2} \Big(\frac{-q}{k^2 z}\Big)^{s_3} dz \nonumber \\ = & B^{s_1} {(\frac{-q}{k^2})}^{s_3} \frac{\Gamma(1+\frac{s_1}{a}+s_2)\Gamma(-1-\frac{s_1}{a}-s_2+s_3)}{\Gamma(s_3)}
\end{flalign}
\begin{flalign}\label{inner_integral_BER_1}
	I_2 =& \int_{1}^{\infty}(z-1)^{-\frac{\phi}{2}} (B(z-1))^{\frac{s_1}{a}} {(z-1)}^{s_2} \Big(\frac{-q}{k^2 z}\Big)^{s_3} dz\nonumber \\ = & B^{s_1} {(\frac{-q}{k^2})}^{s_3} \frac{\Gamma(1+\frac{s_1}{a}+s_2-\frac{\phi}{2})\Gamma(-1+\frac{\phi}{2}-\frac{s_1}{a}-s_2+s_3)}{\Gamma(s_3)}
\end{flalign}

Substituting the inner integrals \eqref{inner_integral_BER_1} and \eqref{inner_integral_BER_2} into \eqref{meiger_def_BER_1}, changing $S_3$ $ \to $ $-S_3$, and utilizing the definition of trivariate Fox's H-function \cite{Mathai_2010}, we get the average BER of the considered system as
\begin{flalign} \label{eq:ber_fox}
	\bar{P}_e =& \frac{\zeta^{\frac{\phi}{\alpha}} q^p}{{\gamma_0}^\frac{\phi}{2} k^{\phi} 2\Gamma(p)} \Bigg[H_{2,0:1,2:1,1:2,0}^{0,2:1,1:1,1:0,2} \Bigg(\frac{\zeta}{\gamma_0^{\frac{\alpha}{2}}k^{\alpha}},1,-\frac{k^2}{q} \Bigg|\begin{matrix} ~D_7~ \\ ~D_8~  \end{matrix} \Bigg)  \nonumber \\ + &  H_{2,0:1,2:1,1:2,0}^{0,2:2,0:1,1:0,2} \Bigg(\frac{\zeta}{{\gamma_0}^\frac{\alpha}{2} k^{\alpha}},1,-\frac{k^2}{q} \Bigg|\begin{matrix} ~D_9~  \\ ~D_{10}~   \end{matrix} \Bigg)  \Bigg]
\end{flalign}
where $D_7 = \bigl\{(0;\frac{\alpha}{2},1,0),(2;\frac{\alpha}{2},1,1)\bigr\}:\bigl\{(1,1)\bigr\}:\bigl\{(0,1)\bigr\}:\bigl\{(1,1);(0,1)\bigr\}$, $D_8 = {\bigl\{-\bigr\}:\bigl\{(\mu,1);(0,1)\bigr\}:\bigl\{(4-p,1)\bigr\}:\bigl\{-\bigr\}}$, $D_9 = \bigl\{(\frac{\phi}{2};\frac{\alpha}{2},1,0),(2-\frac{\phi}{2};\frac{\alpha}{2},1,1)\bigr\}:\bigl\{(1,1)\bigr\}:\bigl\{(0,1)\bigr\}:\bigl\{(1,1);(0,1)\bigr\}$ and $D_{10} = \bigl\{-\bigr\}:\bigl\{(C,1);(0,1)\bigr\}:\bigl\{(4-p,1)\bigr\}:\bigl\{-\bigr\}$.

It is becoming common in the research fraternity to develop performance analysis using Fox's H representation due to the availability of efficient computational routines for the function \cite{Alhennawi_2016_mv_FoxH} and elegant asymptotic expression in terms of the more straightforward Gamma function. 

To derive the asymptotic expression of the average BER in high SNR region, we compute the residues of \eqref{eq:ber_fox} at the poles $s_1=\mu,0$, $s_2=s_3=0$ to get
\begin{flalign}
	\bar{P}_e^{\infty} &= \frac{q^p}{2\Gamma(p)} \frac{\psi \zeta^\frac{-\phi}{\alpha}}{\phi} \bigg(\frac{\zeta}{\gamma_0^{\frac{\alpha}{2}}k^{\alpha}}\bigg)^\mu -  \frac{\zeta^{\frac{\phi}{\alpha}} q^p}{{\gamma_0}^\frac{\phi}{2} k^{\phi} 2\Gamma(p)} B^\mu \Gamma(1-\frac{\phi}{2}) \nonumber \\ \times  & \Gamma(-1+\frac{\phi}{2})  -  3 \frac{\zeta^{\frac{\phi}{\alpha}} q^p}{{\gamma_0}^\frac{\phi}{2} k^{\phi} 2\Gamma(p)} \Gamma(1-\frac{\phi}{2}) \Gamma(-1+\frac{\phi}{2})
\end{flalign}

Observing the exponents of $\gamma_0$, the diversity order for average BER is obtained as $\min\Bigr\{\frac{\alpha\mu}{2}, \frac{\phi}{2}\Bigr\}$, which is the same as derived  for the outage probability.

\subsection{Ergodic Capacity}
 Ergodic capacity provides an estimate of the data rate transmission over fading channels. Thus, we use \eqref{eq:thi_snr} to express the ergodic capacity of the THz wireless system as
    \begin{eqnarray}\label{capacity_1}
    \bar\eta = \int_{0}^{\frac{1}{k^2}} \log_2(1+\gamma) {{f_\gamma(\gamma)}} d\gamma
    \end{eqnarray}
 Substituting  \eqref{simplified_pdf} in \eqref{capacity_1} with Meijer's G equivalent of $ \log_2(1+\gamma) $ and the substitution $({\frac{1}{\gamma k^2} -1 })^{\frac{\alpha}{2}} = t$, \eqref{capacity_1} can be rewritten as
    \begin{flalign}\label{meiger_def_integral_capacity}
    	\bar{\eta} =& \frac{-a\psi}{ 2\gamma_0^\frac{\phi}{2} k^{\phi+2}}  \frac{1}{(2\pi i)^3} \int_{L_1}^{} \int_{L_2}^{}\int_{L_3}{} \frac{\Gamma(C+s_1)\Gamma(0+s_1)}{\Gamma(1+s_1)} {(B)}^{-s_1} \nonumber \\ \times &  \Gamma(s_2) \Gamma(2-s_2) \frac{ \Gamma(1+s_3)\Gamma {(-s_3)}^2 }{\Gamma(1-s_3)} ds_1 ds_2 ds_3  \times  I_3
    \end{flalign} 
where $I_3$ is given by \cite[3.194.3]{Gradshteyn}    
    \begin{flalign}\label{final_meiger}
    I_3 =& \int_{0}^{\infty}  t^{(-s_1-as_2+\frac{-\phi}{2}+\alpha-1)} \bigg(\frac{1}{k^2 (t^a+1)}\bigg)^{(-s_3-1)} dt \nonumber \\ = & k^{2(1+s_3)} \frac{\Gamma(\frac{1-\frac{\phi}{2} +\alpha-1-s_1-as_2}{a}) \Gamma(\frac{-1-\frac{\phi}{2} +\alpha-1-s_1-as_2-a-as_3}{a})}{a\Gamma(-1-s_3)}
    \end{flalign}
    
   Substituting $I_3$ in \eqref{meiger_def_integral_capacity}, and applying the definition of trivariate Fox's H-function, an exact expression for the ergodic capacity is given by
   	\begin{eqnarray} \label{fox_H_capacity}
   	\bar\eta = \frac{\psi}{8\gamma_0^\frac{\phi}{2} k^{\phi+2}} H_{2,0:1,2:1,1:2,3}^{0,2:2,0:1,1:1,2} \Bigg(\frac{{\gamma_0}^\frac{\alpha}{2} k^{\alpha}}{\zeta},1,\frac{1}{k^2} \Bigg|\begin{matrix} ~ D_5~ \\ ~D_6~  \end{matrix} \Bigg)  
   \end{eqnarray}
   where $D_5 = \bigl\{(1-\frac{1-\frac{\phi}{2} + \alpha-a}{a};\frac{\alpha}{2},1,0),(1-\frac{-2-\frac{\phi}{2}+\alpha - a}{a};\frac{\alpha}{2},1,1)\bigr\}:\bigl\{(1,1)\bigr\}:\bigl\{(1,1)\bigr\}:\bigl\{(1,1);(1,1) \bigr\} $, $D_6 = \bigl\{-\bigr\}:\bigl\{(C,1),(0,1)\bigr\}:\bigl\{(2,1)\bigr\}:\bigl\{(1,1),(0,1),(2,1)\bigr\} $.

\begin{figure*}[t]
	\centering
	\subfigure[Outage probability with $\alpha=2$.]{\includegraphics[scale=0.31]{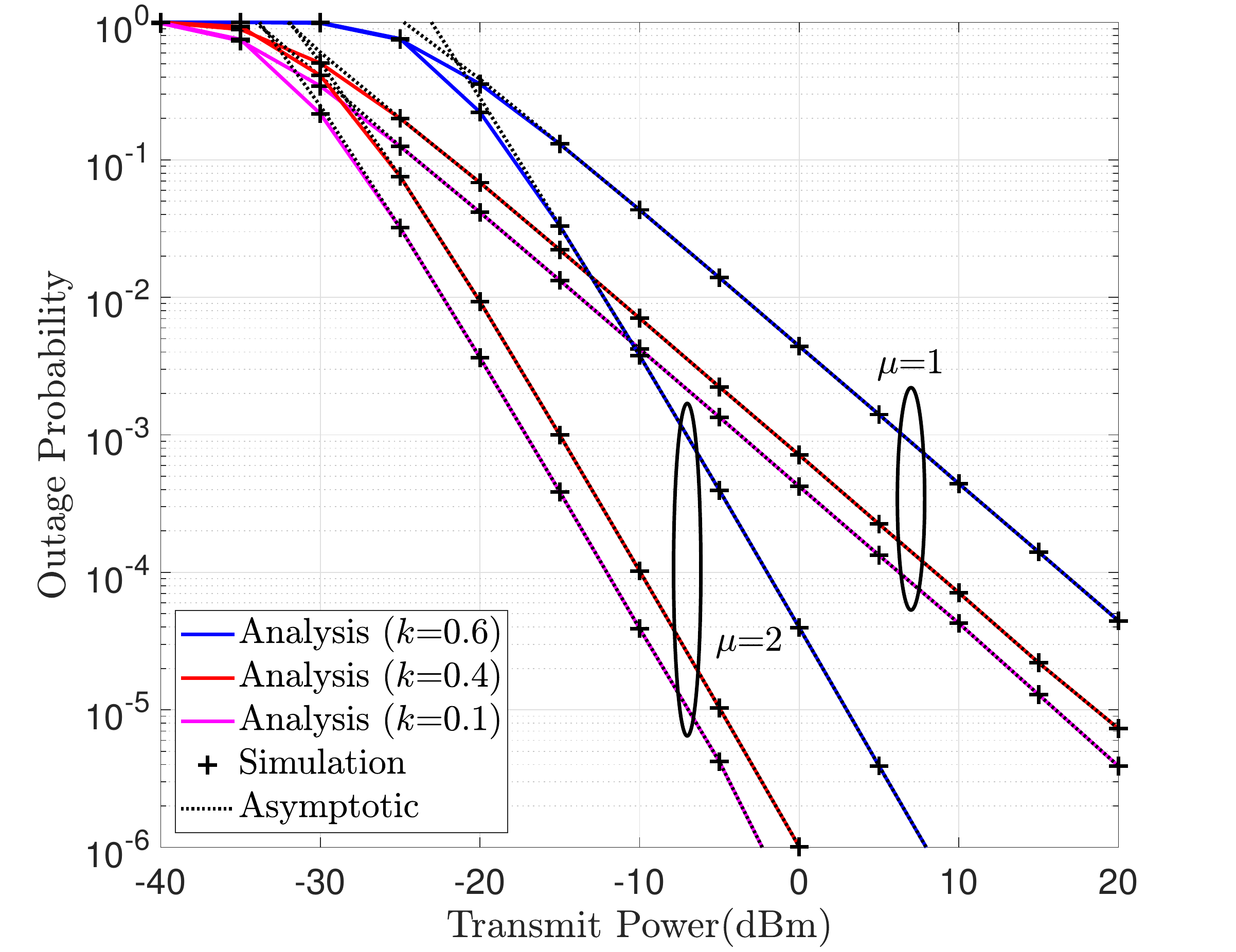}} 
	\subfigure[Average BER with $\mu=1$.]{\includegraphics[scale=0.31]{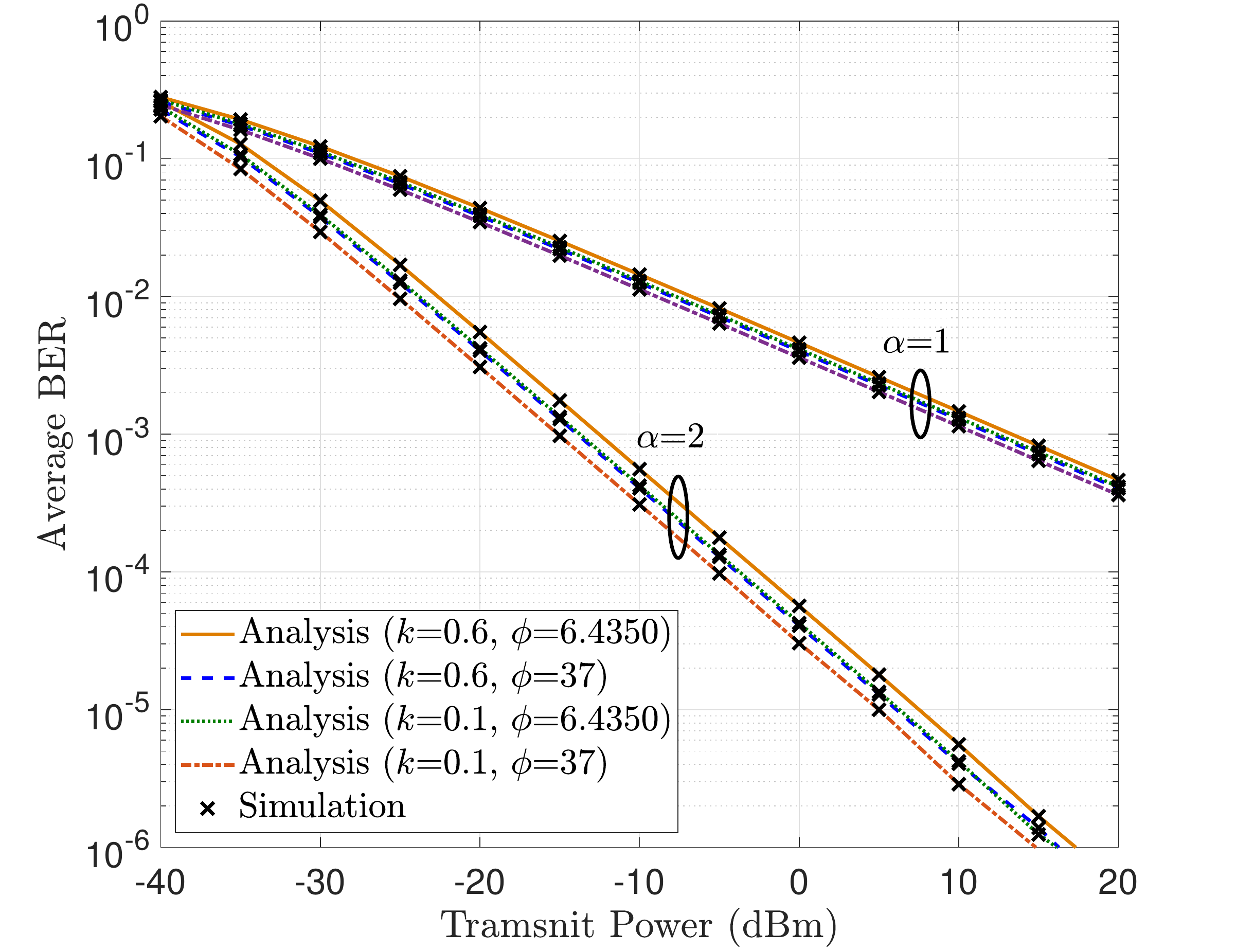}} 
	\caption{Outage probability and average BER performance of the THz wireless system with  THI.}
	\label{fig:outage_ber}	
\end{figure*}

\subsection{$n_{}$-th moment of SNR}
The $n_{}$-th moment of SNR can provide various statistical measures such as average SNR, the variance of SNR, and the amount of fading (AoF). It is defined as
\begin{eqnarray}\label{nth_moment}
	\bar{\gamma}^{(n)} = \int_{0}^{\frac{1}{k^2}} \gamma^{n}  {{f_\gamma(\gamma)}} d\gamma
\end{eqnarray}

Using the PDF of \eqref{simplified_pdf} in \eqref{nth_moment} with two consecutive  substitutions $({\frac{1}{\gamma k^2}-1 })^{\frac{\alpha}{2}} = t$ and $t = \frac{1}{(1-z)^{\frac{1}{a}}}$, we get	
\begin{flalign} \label{eq:derive1}
	\bar{\gamma}^{(n)} =& \frac{-\psi}{8\gamma_0^\frac{\phi}{2} k^{\phi+2n}}\int_{-\infty}^{1} {(-z+1)}^{\frac{\phi}{2}}  \Gamma\bigg(C,\frac{B}{(1-z)^{\frac{1}{a}}}\bigg) \nonumber \\ \times & \frac{1}{(\frac{-z}{2}+1)^{n+1}} dz
\end{flalign}

Further, expressing incomplete Gamma function  and $ \frac{1}{(\frac{-z}{2}+1)^{n+1}} $ using Meijer's G-functions, \eqref{eq:derive1} can be represented as	
\begin{flalign}\label{integral_final}
	\bar{\gamma}^{(n)} =& {\frac{-\psi}{8\gamma_0^\frac{\phi}{2} k^{\phi+2n}}} \int_{-\infty}^{1}  {(-z+1)}^{\frac{\phi}{2}}  G_{1,2}^{2,0} \Bigg(\frac{B}{(1-z)^{\frac{1}{a}}} \Bigg| \begin{matrix} 1 \\ C,0  \end{matrix} \Bigg) \nonumber \\ \times & G_{1,1}^{1,1} \Bigg( \frac{-z}{2} \Bigg| \begin{matrix} 1-n \\ 0 \end{matrix} \Bigg) dt	
\end{flalign}
In the literature, no identity is available for the product of two Meijer's G-functions of the form in \eqref{integral_final} with the given integration limits. Thus, we employ the Mellin–Barnes type integral form of Meijer's G-function and change the order of integrals to express \eqref{integral_final} as 
\begin{flalign} \label{meiger_def_integral}
	\bar{\gamma}^{(n)} =& \frac{-\psi}{ 8\gamma_0^\frac{\phi}{2} k^{\phi+2n} \Gamma(n)} \frac{1}{(2\pi i)^2}  \int_{L_1}^{} \int_{L_2}^{} \frac{\Gamma(C-s_1)\Gamma(-s_1)}{\Gamma(1-s_1)} \nonumber \\ \times & (B)^{s_1} \Gamma(s_2) \Gamma(n+1-s_2)\Big(\frac{-1}{2}\Big)^{-s_2}  ds_1 ds_2 \times  I_4
\end{flalign}
where $I_4$  is represented using\cite[3.191.1,2]{Gradshteyn} as	
\begin{flalign}
	I_4 =& \int_{-\infty}^{1} (1-z)^{\frac{\phi}{2}}  \bigg(\frac{1}{(1-z)^{\frac{1}{a}}}\bigg)^{s_1} z^{-s_2} dz \nonumber \\ = &  2^{s_2} \Gamma(1-s_2) \frac{\Gamma(-1-\frac{\phi}{2}+\frac{s_1}{a}+s_2)\Gamma(1+s_2)}{\Gamma(\frac{s_1-b}{a})} \nonumber \\ + & (-2)^{s_2} \Gamma(1-s_2) \frac{\Gamma(1+\frac{\phi}{2}-\frac{s_1}{a})}{\Gamma(\frac{s_1-b+a(2-s_2)}{a})}
\end{flalign}

Substituting $I_4$ in \eqref{meiger_def_integral}, and applying the definition of bivariate Fox's H-function \cite{Mittal_1972}, the $n$-th moment of the SNR is given by 
\begin{flalign} \label{eq:nth_moment}
	\bar\gamma^{(n)} = & \frac{-\psi}{8\gamma_0^\frac{\phi}{2} k^{\phi+2n} \Gamma(n)} \Bigg[H_{1,0:1,3;1,2}^{0,1:2,0;1,2} \Bigg(\frac{\zeta}{{\gamma_0}^\frac{\alpha}{2} k^{\alpha}},2 \Bigg|\begin{matrix} ~D_1~ \\ ~D_2~ \end{matrix} \Bigg) \nonumber \\ + &  H_{1,0:0,3;1,2}^{0,1:3,0;1,2} \Bigg(\frac{\zeta}{{\gamma_0}^\frac{\alpha}{2} k^{\alpha}},-2 \Bigg|\begin{matrix} ~D_1~ \\ ~D_2~ \end{matrix} \Bigg)\Bigg] 
\end{flalign}
where $D_1 = \big\{(2+\frac{\phi}{2};\frac{\alpha}{2},1)\big\}:\bigl\{(1,1)\bigr\}:\bigl\{(1,)\bigr\}$ and $D_2=\big\{-\big\}:\big\{(C,1),(0,1),(1+\frac{\phi}{2},\frac{\alpha}{2})\big\}:\bigl\{(1,1);(n+1,1)\bigr\}$, $D_3=\bigl\{(2+\frac{\phi}{2};\frac{\alpha}{2},1)\bigr\}:\big\{-\big\}:\bigl\{(1,1)\bigr\}  $ and $D_4=\big\{-\big\}:\bigl\{(C,1);(0,1);(1+\frac{\phi}{2},\frac{\alpha}{2})\bigr\}:\bigl\{(1,1);(n+1,1)\bigr\}$.

\begin{figure*}[t]
	\centering
	\subfigure[Loss in average SNR with $\alpha=1$ and $d=100 \mbox{m}$.]{\includegraphics[scale=0.31]{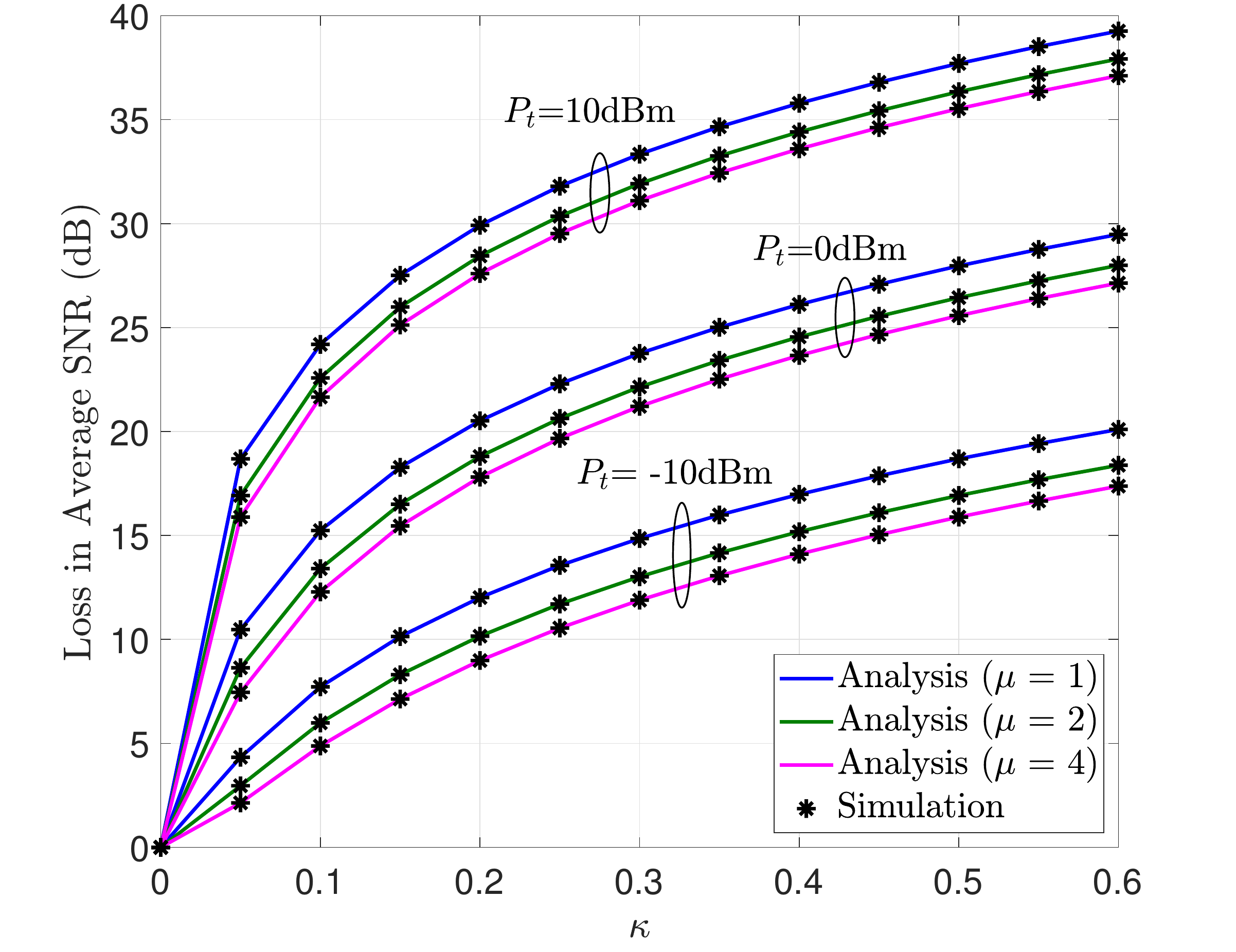}} 
	\subfigure[Loss in ergodic capacity with $\alpha=2, \mu = 2$, and $P= 0$\mbox{dBm}. ]{\includegraphics[scale=0.31]{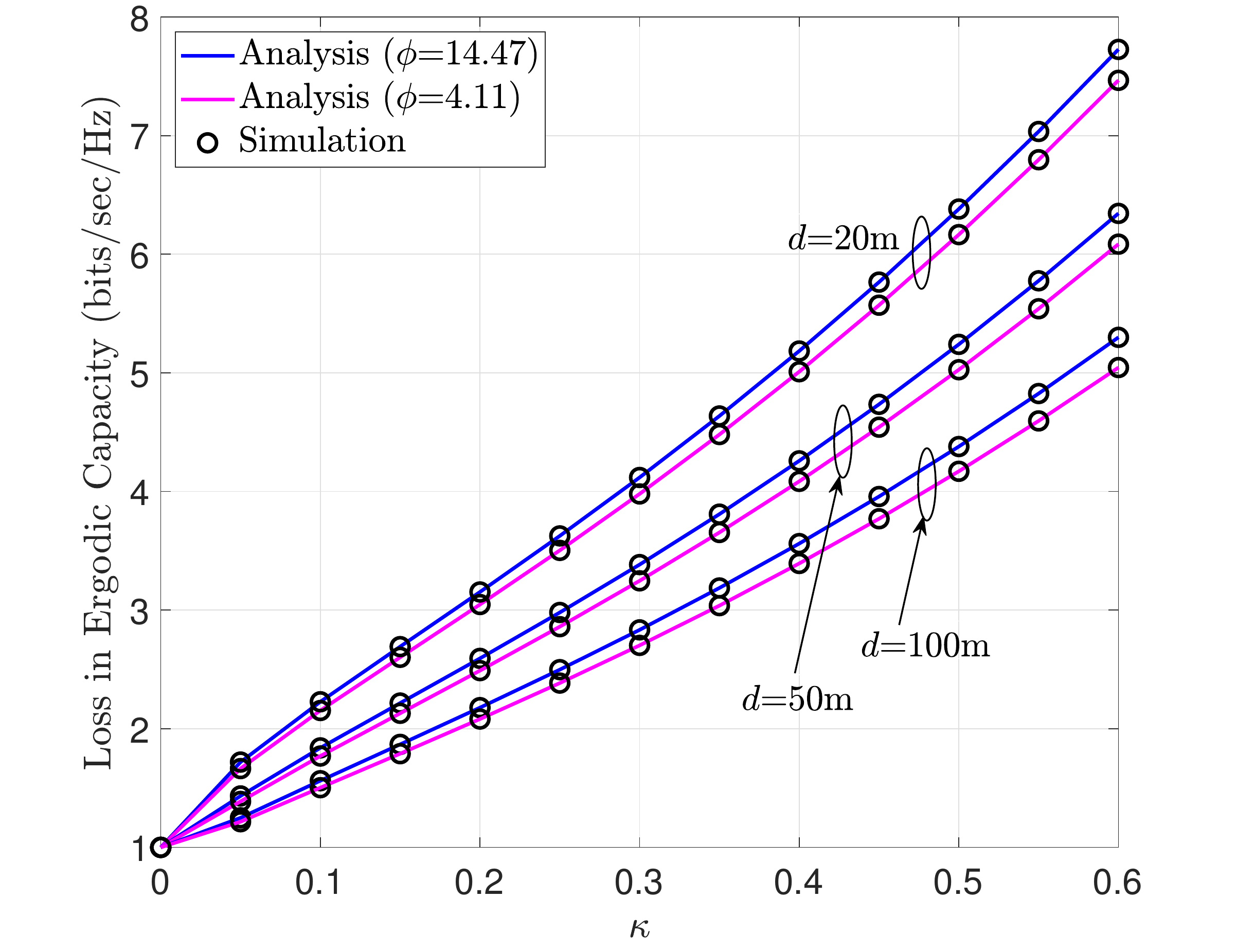}} 
	\caption{Loss in average SNR and ergodic capacity with the THI parameter $k$.}
	\label{fig:snr_rate}	
\end{figure*}

\section{Numerical and Simulation Results}
In this section, we demonstrate the impact of THI on the THz performance using Monte Carlo simulations and numerical evaluations of the derived analytical results. We simulate  $h_l = \frac{c\sqrt{G_{t}G_{r}}}{4\pi f_c l} \exp(-\frac{1}{2}\kappa l)$ at a carrier frequency $f_c=275$ \mbox{GHz}, antenna  gains $G_t=G_r=55$ \mbox{dBi}, distance range $l=$ $20$\mbox{m}-$100$\mbox{m}, and absorption coefficient $\kappa=$ \cite{Boulogeorgos_Analytical} \cite{Pranay_2021_TVT}. We use the MATLAB function \emph{gamrnd($\cdot$)} to simulate $\alpha$-$\mu$ channel for different fading parameters. To numerically evaluate the derived analytical expressions, we use the software package from  \cite{Alhennawi_2016_mv_FoxH} to compute the underlying Fox's H-functions.

We demonstrate the outage probability and average BER performance of the THz link by varying the THI parameter $k$ for different channel parameters at a link distance $d=50$ \mbox{m} in Fig. \ref{fig:outage_ber}. We fix $\alpha=2$ and pointing error parameter $\phi = 14.41$ for the outage probability, as depicted in Fig. \ref{fig:outage_ber}(a). The parameter $k$ has a significant effect on the outage probability. It can be seen from Fig. \ref{fig:outage_ber}(a) that the outage probability increases by almost $10$ times when $k$ is increased from $0.1$ to $0.6$. Moreover, an increase in multi-path clustering parameter $\mu$ improves the channel condition, resulting an improved outage performance, as demonstrated in previous literature \cite{Boulogeorgos_Analytical} \cite{Pranay_2021_TVT}. Also, the slope of the outage probability does not change with $k$ but  with $\mu$ validating our derived diversity order. In Fig. \ref{fig:outage_ber}(b), we further substantiate the effect of the THI on the THz performance by simulating the average BER with BPSK modulation using $p=0.5$ and $q=1$. Here, we fix the fading channel at $\mu=1$ and vary other parameters such $\alpha$, $k$, and pointing error parameter $\phi$. Although the average BER increases with an increase in $k$, the effect of the THI on the average BER is marginal when the fading channel is highly non-linear. However, as the channel becomes linear, the effect of the THI on the average BER performance becomes significant. The average BER shows the expected trend with higher $\phi$ (lower pointing errors) and higher $\alpha$ (linear channel). Most importantly, the slope of the plots confirms the derived diversity order of the system: independent of $k$ and depends on the parameter $\alpha$.

In Fig. \ref{fig:snr_rate}, we plot loss in the average SNR (by evaluating the ratio between average SNR without THI and average SNR with THI) and loss in the ergodic capacity (by evaluating the ratio between ergodic capacity without THI and ergodic capacity with THI) versus parameter $k$ for different channel and system parameters. In Fig. \ref{fig:snr_rate}(a), we illustrate the loss in the average SNR at different transmit powers and fading parameter $\mu$ at a link distance $d=100$\mbox{m} by fixing $\alpha=1$, $\phi=14.41$. The loss in the average SNR with the THI is higher at a higher transmit power, advocating limiting the transmit power to minimize the hardware impairment. It can be seen that the loss in the average SNR decreases when the channel conditions improve at  higher values of $\mu$. Finally, Fig. \ref{fig:snr_rate}(b) shows the impact of link distance on the ergodic capacity of the THz transmission with the THI parameter $k$. It can be observed that there is a significant loss in the ergodic capacity when the link distance is lower: almost $2.5$ bits/sec/Hz loss in the ergodic capacity if the link distance is reduced from $100$\mbox{m} to $20$\mbox {m} at $k=0.6$. Fig. \ref{fig:snr_rate}(b) also shows that the impact of link distance on the ergodic capacity with the THI is higher when the parameter $k$ is high.
\section{Conclusion}
We developed exact analytical expressions for the outage probability, average BER, ergodic capacity, and the $n$-th moment of SNR for a THz transmission with the THI, channel fading, and pointing errors. We also derived asymptotic expressions for the outage probability and average BER demonstrating that the diversity order of the system is independent of the THI parameter but depends on the channel fading and pointing errors parameters. We presented numerical and simulation results to demonstrate that the impact of hardware impairment is more significant for  THz transmissions operating at a higher value of average SNR.

\bibliographystyle{ieeetran}
\bibliography{THz_hardware}	

\begin{thebibliography}{10}
\providecommand{\url}[1]{#1}
\csname url@samestyle\endcsname
\providecommand{\newblock}{\relax}
\providecommand{\bibinfo}[2]{#2}
\providecommand{\BIBentrySTDinterwordspacing}{\spaceskip=0pt\relax}
\providecommand{\BIBentryALTinterwordstretchfactor}{4}
\providecommand{\BIBentryALTinterwordspacing}{\spaceskip=\fontdimen2\font plus
\BIBentryALTinterwordstretchfactor\fontdimen3\font minus
  \fontdimen4\font\relax}
\providecommand{\BIBforeignlanguage}[2]{{%
\expandafter\ifx\csname l@#1\endcsname\relax
\typeout{** WARNING: IEEEtran.bst: No hyphenation pattern has been}%
\typeout{** loaded for the language `#1'. Using the pattern for}%
\typeout{** the default language instead.}%
\else
\language=\csname l@#1\endcsname
\fi
#2}}
\providecommand{\BIBdecl}{\relax}
\BIBdecl

\bibitem{Akyildiz_2020}
I.~F. Akyildiz \emph{et~al.}, ``{6G} and beyond: The future of wireless
  communications systems,'' \emph{IEEE Access}, vol.~8, pp. 133\,995--134\,030,
  2020.

\bibitem{Dang_2020_nature}
S.~Dang \emph{et~al.}, ``What should {6G} be?'' \emph{Nature Electron}, no.~3,
  p. 20–29, 2020.

\bibitem{Schenk2008}
T.~Schenk, \emph{RF Imperfections in High-rate Wireless Systems Impact and
  Digital Compensation}.\hskip 1em plus 0.5em minus 0.4em\relax Dordrecht, The
  Netherlands:Springer, 2008.

\bibitem{Boulogeorgos_Error}
A.~A. {Boulogeorgos} and A.~{Alexiou}, ``Error analysis of mixed {THz-RF}
  wireless systems,'' \emph{IEEE Commun. Lett.}, vol.~24, no.~2, pp. 277--281,
  2020.

\bibitem{Boulogeorgos_2020_THz_THz}
A.-A.~A. Boulogeorgos and A.~Alexiou, ``Outage probability analysis of {THz}
  relaying systems,'' in \emph{2020 IEEE 31st Annu. Int. Symp. on Personal,
  Indoor and Mobile Radio Commun.}, 2020, pp. 1--7.

\bibitem{Pranay_2021_TVT}
P.~Bhardwaj and S.~M. Zafaruddin, ``Performance of dual-hop relaying for
  {THz-RF} wireless link over asymmetrical $\alpha$-$\mu$ fading,'' \emph{IEEE
  Trans. Veh. Technol.}, vol.~70, no.~10, pp. 10\,031--10\,047, 2021.

\bibitem{Li_2021_THz_AF}
S.~Li and L.~Yang, ``Performance analysis of dual-hop {THz} transmission
  systems over $\alpha$-$\mu $ fading channels with pointing errors,''
  \emph{IEEE Internet of Things J.}, pp. 1--1, 2021.

\bibitem{Bhardwaj2022_multihop}
P.~Bhardwaj and S.~M. Zafaruddin, ``On the performance of multihop {THz}
  wireless system over mixed channel fading with shadowing and antenna
  misalignment,'' \emph{IEEE Trans. Commun.}, vol.~70, no.~11, pp. 7748--7763,
  2022.

\bibitem{Boulogeorgos_Analytical}
A.~A. {Boulogeorgos} and A.~{Alexiou}, ``Analytical performance assessment of
  {THz} wireless systems,'' \emph{IEEE Access}, vol.~7, pp. 11\,436--11\,453,
  2019.

\bibitem{Boulogeorgos2022_rain}
A.-A.~A. Boulogeorgos \emph{et~al.}, ``On the joint effect of rain and beam
  misalignment in terahertz wireless systems,'' \emph{IEEE Access}, vol.~10,
  pp. 58\,997--59\,012, 2022.

\bibitem{Du2022_FTR_THz_RIS}
H.~Du \emph{et~al.}, ``Performance and optimization of reconfigurable
  intelligent surface aided {THz} communications,'' \emph{IEEE Trans. Commun.},
  vol.~70, no.~5, pp. 3575--3593, 2022.

\bibitem{Chapala2021_letters}
V.~K. Chapala and S.~M. Zafaruddin, ``Exact analysis of {RIS}-aided {THz}
  wireless systems over $\alpha$-$\mu$ fading with pointing errors,''
  \emph{IEEE Commun. Lett.}, vol.~25, no.~11, pp. 3508--3512, 2021.

\bibitem{Bjornson2013_HW_dualhop}
E.~Bjornson \emph{et~al.}, ``A new look at dual-hop relaying: Performance
  limits with hardware impairments,'' \emph{IEEE Trans. Commun.}, vol.~61,
  no.~11, pp. 4512--4525, 2013.

\bibitem{Balti2018_HW_FSORF}
E.~Balti \emph{et~al.}, ``Aggregate hardware impairments over mixed {RF/FSO}
  relaying systems with outdated {CSI},'' \emph{IEEE Trans. Commun.}, vol.~66,
  no.~3, pp. 1110--1123, 2018.

\bibitem{Mathai_2010}
A.~M.{ Mathai} \emph{et~al.}, \emph{The {H}-Function: Theory and
  Applications}.\hskip 1em plus 0.5em minus 0.4em\relax Springer, New York, NY,
  2010.

\bibitem{Papasotiriou2021_scientific_report}
E.~Papasotiriou \emph{et~al.}, ``An experimentally validated fading model for
  {THz} wireless systems,'' \emph{Scientific report}, vol.~11, 2021.

\bibitem{Farid2007}
A.~A. {Farid} and S.~{Hranilovic}, ``Outage capacity optimization for
  free-space optical links with pointing errors,'' \emph{J. Lightw. Technol.},
  vol.~25, no.~7, pp. 1702--1710, 2007.

\bibitem{Dabiri2022}
M.~T. Dabiri and M.~Hasna, ``Pointing error modeling of {mmWave} to {THz}
  high-directional antenna arrays,'' \emph{IEEE Wireless Commun. Lett.},
  vol.~11, no.~11, pp. 2435--2439, 2022.

\bibitem{papoulis_2002}
{A. Papoulis} and {S. Pillai}, \emph{Probability, Random Variables, and
  Stochastic Processes}.\hskip 1em plus 0.5em minus 0.4em\relax McGraw Hill,
  Boston, Fourth Edition, 2002.

\bibitem{Jameson2016}
G.~J.~O. {Jameson}, ``The incomplete gamma functions,'' \emph{The Mathematical
  Gazette}, vol. 100, Iss. 548, pp. 298--306, Jul 2016.

\bibitem{Gradshteyn}
I.~S. {Gradshteyn} and I.~M. {Ryzhik }, \emph{Table of Integrals, Series, and
  Products}.\hskip 1em plus 0.5em minus 0.4em\relax Academic press, San Diego,
  CA, 6th edition, 2000.

\bibitem{Alhennawi_2016_mv_FoxH}
H.~R. Alhennawi \emph{et~al.}, ``Closed-form exact and asymptotic expressions
  for the symbol error rate and capacity of the {H-Function} fading channel,''
  \emph{IEEE Trans. Veh. Technol.}, vol.~65, no.~4, pp. 1957--1974, 2016.

\bibitem{Mittal_1972}
P.~{Mittal} and K.~{Gupta}, ``An integral involving generalized function of two
  variables,'' \emph{Proc. Indian Acad. Sci.}, vol.~75, no.~9, pp. 117--123,
  1972.

\end{thebibliography}

\end{document}